**Journal title**

*Genome Biology*

**Article type**

Review

**Paper title**

Metagenomics for clinical diagnostics: technologies and informatics

**Authors and affiliations**

Caitlin Loeffler[1,2], Keylie M. Gibson[3], Lana S. Martin[2], Yutong Chang[4], Jeremy Rotman[2], Ian V. Toma[5], Christopher E. Mason[6,7,8], Eleazar Eskin[1,9,10], Joseph P. Zackular[11], Keith A. Crandall[3,12], David Koslicki*[13,14,15], Serghei Mangul*[2]

[1]Department of Computational Medicine, David Geffen School of Medicine, University of California Los Angeles, Los Angeles, CA, USA

[2]Department of Clinical Pharmacy, School of Pharmacy, University of Southern California, Los Angeles, CA, USA

[3]Computational Biology Institute, Milken Institute School of Public Health, George Washington University, Washington, DC, USA

[4]Department of Pharmacology and Pharmaceutical Sciences, School of Pharmacy, University of Southern California, Los Angeles, CA, USA

[5]Clinical Research & Leadership, School of Medicine and Health Sciences, George Washington University, Washington, DC, USA

[6]Department of Physiology and Biophysics, Weill Cornell Medicine, New York, NY, USA

[7]The HRH Prince Alwaleed Bin Talal Bin Abdulaziz Alsaud Institute for Computational Biomedicine, Weill Cornell Medicine, New York, NY, USA

[8]The WorldQuant Initiative for Quantitative Prediction, Weill Cornell Medicine, New York, NY, USA

[9]Department of Computer Science, University of California Los Angeles, 580 Portola Plaza, Los Angeles, CA 90095, USA

[10]Department of Human Genetics, David Geffen School of Medicine at UCLA, 695 Charles E. Young Drive South, Box 708822, Los Angeles, CA, 90095, USA

[11]Department of Pathology and Laboratory Medicine, University of Pennsylvania and Children's Hospital of Philadelphia, Philadelphia, PA, USA

[12]Department of Biostatistics & Bioinformatics, George Washington University, Washington, DC, USA

[13]Department of Computer Science and Engineering, The Pennsylvania State University, University Park, PA, USA

[14]Department of Biology, The Pennsylvania State University, University Park, PA, USA

[15]Huck Institutes of the Life Sciences, The Pennsylvania State University, University Park, PA, USA

*Co-senior authors.

**Abstract**

The human-associated microbiome is closely tied to human health and is of substantial clinical interest. Metagenomics-based tools are emerging for clinical diagnostics, tracking the spread of diseases, and surveillance of potential pathogens. In some cases, these tools are overcoming limitations of traditional clinical approaches. Metagenomics has limitations barring the tools from clinical validation. Once these hurdles are overcome, clinical metagenomics will inform doctors of the best, targeted treatment for their patients and provide early detection of disease. Here we present an overview of metagenomics methods with a discussion of computational challenges and limitations.

**Introduction**

A microbiome is the microscopic ecosystem of a given habitat composed of communities of microbes (bacteria, archaea, fungi, protists, parasites, viruses) that simultaneously exist and interact both inside and outside a microbiome [1]. Nearly all environments on Earth are inhabited by dynamic microbial communities, including the environments on, around, and inside humans. The content of these human-associated microbiomes and their relationship to changes in host diet, environment, and health have become the subject of an increasing number of studies. Omics-based methods have emerged as the most popular tools used to elucidate the genetic content of humans and microbes and determine microbe abundance within human-associated microbiomes.

Omics studies include, but are not limited to, epigenetics, transcriptomics, and genomics. Epigenomics focuses on gene expression by studying the chemical markers, such as methyl groups, placed on the DNA or on the histones wrapped in DNA [2,3]. Epigenomics also includes the study of how environmental factors can change these chemical markers [2,3]. Oncology currently has FDA-approved translational epigenetic therapies – termed 'epidrugs' – that are integrated into medical guidelines and practices for cancer detection in humans [4]. Alternatively, transcriptomics focuses on messenger RNA. Quantifying specific sets of detected messenger RNA allows researchers to study gene expression within a single cell and identify genes that are active in specific tissues [3,5]. While not yet approved for clinical use, a variety of clinical trials suggest transcriptomics have strong potential to be useful in patient care settings [6], and transcriptomics are already used in ENCODE (Encyclopedia of the regulatory elements) and TCGA (The Cancer Genome Atlas) [7].

Metagenomics is a class of genomics, the study of the genome sequence [3], that considers the entirety of all genome sequences (i.e., the metagenome) from organisms within a sampled microbiome. In this review, we define clinical metagenomics as the characterization of a patient's microbiome, through analysis of the metagenome, for the purpose of diagnosing disease, tailoring treatment, and improving quality of life. The clinical metagenomics workflow excludes culturing organisms and targeted amplification steps, and involves sequencing and analyzing only the DNA directly gathered from a patient's tissue, swab, fluid, or solid sample.

Metagenomics relies on high-throughput genomics technologies, which have become increasingly reliable and affordable within the past decade, leading to the exponential rise in metagenomic studies within the fields of microbiology and genomics. These Next Generation Sequencing (NGS) technologies offer increased DNA sequencing depth and throughput which has drastically increased the amount of sequencing data produced at reduced cost compared to first-generation sequencing technologies. In response to overwhelming data increases, bioinformatics software is continually developed to meet the analytical need of researchers. With a significant reduction in cost of sequencing technologies and increase in analytical power of computational software, there is enormous potential for leveraging microbiome data for development of novel diagnostic and therapeutic tools in the clinical setting.

Identification of microbes from diverse sample types can help determine their impact on the microbiome, metagenomics. Together with associative and causative studies, metagenomics has substantial potential as a tool for improving the diagnosis and treatment of disease [8]. Medical applications of metagenomics include, but are not limited to, infection detection, surveillance and tracking of outbreaks, antimicrobial resistance prediction and strain resolution, and disease prognosis. Human biomedical studies focus on microbial diversity across body compartments (e.g., gut, skin, nasal, oral), the niche specialization within those compartments [9], and how these variables influence patient response to treatments [10,11]. Characterization of the human-associated microbiome driven by metagenomics methods and technologies has the potential to be applied as a lifesaving clinical diagnostic tool. Yet, integration of metagenomics as standard clinical practice has yet to be actualized, illustrated by the relatively small number of clinical metagenomics studies published in comparison to metagenomics studies (Figure 1a, 1b).

In this review, we will discuss the advantages of metagenomics techniques in comparison to the current clinical gold standards, as well as the limitations that bar these methods from becoming broadly applied. We outline the complexities and limitations of computational algorithms that handle output from sequencing technologies. An in depth discussion of the specific steps of mapping and report is beyond the scope of this paper; other reviews have covered this topic in depth [12,13]. We reviewed the materials and methods published in 29 clinical metagenomics papers between 2015 and 2019 and briefly discuss the hurdle of clinical validation. Our literature review suggests that metagenomics could be productively applied in clinical labs after clinical validation is achieved by citing relevant studies of metagenomics methods.

**The current state of clinical laboratories**

We define clinical labs as those that diagnose patients with the intent to prevent or treat diseases and determine patient health by processing human samples (e.g., tissue, fecal, blood, swabs) with a validated test. Clinical labs process samples for clinical researchers and practicing doctors; both hospital labs, where your blood is sent for routine testing, and private clinical labs, which receive samples from hospitals for testing, are common. Numerous complex laws guide clinical practice and research. Labs that supply results to doctors for the purpose of informing treatment plans are subject to legal regulation and certification under the Clinical Laboratory Improvement Amendments (CLIA) of 1988 [14,15]. In the United States, tests that are used in the practice and application of medicine on humans generally must be validated and are subject to regulation by both federal and state government agencies. This regulation mainly applies to clinical practice (research is regulated by different laws); while tests and methods for diagnosing certain illnesses may exist, they are not necessarily available to physicians as a diagnostic for patient health care until after a lengthy, rigorous validation process. Current techniques used in validated clinical diagnostic methods include, but are not limited to, culture, DNA amplification by Polymerase Chain Reaction (PCR), and DNA sequencing.

Subsection: *On the use of culture in clinical labs*

Most standard diagnostic tests utilize culture methods – the gold standard for the last 300 years – to identify pathogens or other biomarkers in the human-associated microbiome. Culturing methods require growth of sample microbes in an artificial environment (i.e., culture) in order to identify, quantify, and phenotypically characterize microbial organisms. These artificial, controlled environments use a nutrient-heavy media designed to enrich or grow microbes with specific requirements. Once cultivated, these organisms can be phenotypically characterized using different microbiological and biochemical diagnostic techniques, including microbe morphology, growth, and response to environmental variables (e.g., antibiotics, gel composition) [16]. Culturing methods are widely applied in clinical pathology and microbial research laboratories in order to conduct phenotypic discrimination of bacterial colonies, examination of intracellular pathogenicity, and assess susceptibility to antibiotics [17].

Culturing methods are reliable, reproducible, validated, and widely trusted in both clinical and research labs. However, these methods are not without serious limitations. Culturing procedures are slow, labor intensive, and cannot be scaled across a large number of samples. Growth of microbes in culture precedes the diagnostic step, creating a delay in diagnosis and treatment. Most infectious pathogens take one to two days to grow in culture; other, such as ulcer-causing *Helicobacter pylori*, take five days or more [17,18] or else may be difficult to culture (increasing false negatives). In addition, profiling beyond species exclusively with culturing methods is challenging. For example, with *Mycobacterium tuberculosis* strain information is critical for correct treatment. Moreover, we still know little about the nutrient and environmental requirements of many microbes. Much of the microbial diversity observed in the microbiome has yet to be cultured because pathologists and other researchers are unable to replicate biologically crucial aspects of their host environment in the laboratory setting. This limitation extends beyond bacteria to the culture of viruses, which cannot be cultured without host cells [19]. Indeed, exclusive reliance of clinical testing and microbial research on culture has been a substantial hurdle for the study and identification of uncultured microbes.

A number of these limitations were overcome with the development of DNA sequencing technology and the advent of Polymerase Chain Reaction (PCR). Used together, these technologies allowed researchers to sequence the base pairs in segments of DNA (called reads). Once researchers developed technology capable of analyzing the content of DNA , they were able to sequence the human genome [20,21] and begin viewing the base pair pattern within the DNA as molecular biomarkers for the classification of microorganisms.

Subsection: ***On the use of sequencing and PCR in clinical labs***

A significant a scientific breakthrough, sequencing was first achieved in the mid-1970s with the development of the Sanger sequencing method [22]. Today's sequencing platforms include Illumina, PacBio, and Nanopore – collectively now known as NGS technology. One major limitation of sequencing microbial genomes is that high concentrations of the required pure DNA are not always available, especially if the target microbe cannot be cultured [23] or when the sample metagenome is dominated by DNA from a single organism (e.g., host organism). PCR is often used to provide adequate DNA quantity needed for high quality sequencing and proper coverage of targeted genomes [24]. PCR, developed in the late 1980s by Kary Mullis [25], artificially amplifies specific regions of DNA in order to deepen coverage of those regions for sequencing, perform functional analysis of a specific gene, or detect polymorphisms and point mutations [25–27].

The combination of PCR and sequencing can be applied to both cultured and uncultured microbes. Until development of this combination, pathologists had continually failed to grow microbes using culture. PCR can amplify the amount of DNA in a sample, enabling detection of material even if the amount of microbial DNA available at the outset was incredibly small. PCR and sequencing allowed clinics to diagnose and study infectious pathogens, such as *Mycobacterium leprae* (a bacterium causing leprosy) [28–30], which researchers and pathologists had continually failed to grow and classify using culture methods [31]. The application of these technologies has not eliminated culture methods. A number of culture-based tests are still conducted, including throat (detect strep throat), urine (identify pathogens in urinary tract infections), sputum (detect respiratory infections), blood (detect bacterial or fungal infections), and stool cultures (detect bacteria or parasites) [18].

The continued reliance on culture in the face of advancing technology is partly due to the number of limitations associated with both PCR and sequencing. PCR has four major sources of bias: template switching (leading to the formation of chimera reads) [32], *taq* DNA polymerase sequencing errors [33], amplification bias [34], and stochastic variation [35,36]. PCR can increase contamination as running a large number of PCR cycles may amplify very low abundant microbes that are not actually present in a sample but are introduced from reagents or lab instruments. Sequencing also introduces errors in the reading of base pairs such as insertions, deletions, and substitutions of base pairs [37].

An additional variable in the accuracy of an identified genetic sequences is the quality and comprehensiveness of the reference genome database employed by the bioinformatics data analysis pipeline. Numerous genomic reference databases are available, and differing maintenance practices and standards have resulted in a lack of comprehensiveness. For example, a number of species or genera sequences may be included in one database but absent in another [38]. Inconsistent or incomplete genomic reference databases can cause incorrect and inconsistent detection of microbes from sequence data. The accuracy of data analysis is decreased by errors in sequencing, introduced during PCR or read sequencing steps, and inconsistent genomics reference databases. Effects of biases and limitations are most noticeable when attempting to detect microbes in low abundance.

These limitations are not insurmountable, as proven by the translation of PCR-based sequencing methods to clinical diagnostic labs. Indeed, researchers and engineers are working tirelessly to improve the reliability of PCR and sequencing-based methodologies at each step of the workflow. Special customized genomic reference databases are generated specifically for the application of PCR and sequencing to isolate the genomes of known pathogens for the clinical lab setting. Today, methods for sequencing PCR outputs are applied in certified clinical labs to detect a number of diseases, including influenza [39], multiple meningitis pathogens [40,41], and, recently, COVID-19 [42,43]. Successful applications of these technologies suggest it is realistic to explore applying other methods to clinical diagnostics, specifically the application of metagenomics methods.

**Current approaches and corresponding challenges to metagenomics**

A metagenomics study considers the genes and genomes of all microorganisms from a microbiome (i.e., the metagenome) [1]. DNA composing the metagenome is collected directly from a sample of the studied host environment, which can be collected through biopsies, swabs, or human excretions. These samples can then be stored at either -20 or -80 degrees Celsius to keep DNA or RNA intact, respectively; alternatively, samples can be stored with preservation solutions, such as Zymo DNA/RNA shield [44] or RNAlater Stabilization Solution [45], which open cells and preserve nucleic acids (Figure 2a). Nucleic acid is then extracted, processed, sheared, and sequenced to isolate the base pair patterns (Figure 2b, 2c, 2d, 2e). This process, called metagenomic shotgun sequencing, produces a number of sequences (referred to as reads) that can be assembled into larger contiguous pieces (or directly mapped to reference genomes without assembly) and taxonomically characterized via sequence homology. Generating and analyzing results requires the use of multi-step computational pipelines and algorithms (Figure 2f). In clinical diagnostic settings, computational analysis results are then presented in a clinical report that explains the results for health professionals (Figure 2g, 2h).

As with PCR and culture-based methods, metagenomics methods have inherent limitations. Different aspects of the metagenomics pipeline, such as the sequencing technology or sequencing depth, can drastically change the cost of such methods. Today's computational algorithms for metagenomic analysis also have limitations. Genetic data obtained through metagenomics methods contain an increased number of DNA fragments that must be aligned against hundreds of thousands of microbial genomes, each acquired from inconsistent reference databases, in order to be analytically useful. This increased number of DNA fragments—and the large size of the reference database—poses unique challenges to existing metagenomics methods and the computational resources they depend on [46]. These DNA fragments can be assembled into larger contigs, but the assemblies are not necessarily accurate to the genomes present in a sample, which are instead assembled into fragments and containing chimeras. Additional challenges include the lack of comprehensive gene catalogs, biases in functional profiling, and lack of standardization in publishing raw data. Difficulty integrating meta-omics analysis tools with existing frameworks also limit and challenge the potential use of metagenomics methods.

Provided the sample contains sufficient DNA for processing, metagenomics methods are capable of analyzing the entire collection of genomes and genes from all microbes present in the microbiome [47,48] without relying on culture or PCR amplification. Metagenomics methods are particularly advantageous due to their ability to quantitatively characterize the microbial communities in a sample by assessing the relative abundance of and potential interactions among species. PCR-based techniques are targeted by nature due to the use of primers that bind to specific segments of DNA and engage *taq* DNA polymerase for replication; therefore, microbes whose DNA does not get amplified become lost in analysis. Additionally, metagenomics allows for the characterization of genomic features within the sample and allows for comparison across multiple samples and sample types. However, such methods cannot differentiate between living and dead microbes [49] (this can be achieved through metatranscriptomics for approximately double the cost) and have limited ability to detect microbes that have low DNA representation in a sample. Metagenomics methods can detect the presence of viruses, fungus, yet-to-be-cultured bacteria, and hard-to-culture bacteria given sufficient sequencing depth. The use of complex computational algorithms needed in the analysis of metagenomics methods allow for a deeper genomic analysis of the entire microbiome when compared to culture or PCR and sequencing-based methods alone.

When applying metagenomics methods to clinics, we emphasize that factors relevant for researchers are not necessarily important for clinical diagnostic labs that process samples for physicians. Due to Health Insurance Portability and Accountability Act (HIPAA) regulations that protect patient confidentiality, it is infeasible to publish raw metagenomics data generated by clinical laboratory workflows that process patient samples. Metagenomics researchers, in an effort to increase reproducibility, are making raw sequencing data sourced in studies publicly available. The outcome of a method must be reproducible, and therefore reliable, before application to clinics; however, currently 30% of raw data in clinical metagenomics research remains unpublished (Figure 3). In addition, research sample sizes are expected to be larger in order to produce statistically significant results. Yet, a review of the current clinical metagenomic studies over the last five years (2015-2019) shows fresh and frozen sample sizes varying from 1 to 204 (Supplementary Table 1).

Subsection: *Sample collection and processing limitations and challenges of metagenomics*

There are unique challenges associated with processing different sample types, and current processing techniques have created demand for enrichment protocols designed to remove host DNA. Sources for studying the microbiome within humans come in a variety of forms, each of which carry their own difficulties during the DNA processing that takes place prior to sequencing. In addition, the type of sample collected impacts sample storage and nucleic acid extraction options, which can, in turn, also affect downstream analyses and enrichment protocols. These sources include solids, such as feces; tissues, such as colon, cancer cells, or other biopsy samples; swabs, such as from sampling the skin, oral, nasal, or vaginal epithelial cells; and liquids, such as saliva, breast milk, urine, or blood (Figure 2a). Liquid samples are often frozen directly from sampling at -20 or -80 degrees Celsius, due to the preservation solution diluting the already low amount of metagenomic DNA. In solids, a preservation solution can be used to break cells open and preserve the nucleic acids, then the samples are frozen to allow for extended storage time. Swabs can be stored either way, but an additional factor to take into account is shipment; if samples need to be shipped, especially internationally, the use of a preservation solution is required to inactivate potentially harmful agents.

Following storage, there are many protocols for isolating nucleic acid from samples that can be used to target specific types of microbes and different nucleic acids, all of which should be considered when addressing how to answer the question of interest. Some protocols are optimized for specifically isolating RNA over DNA, viruses over host or bacterial DNA, or are optimized for retrieving nucleic acids from tissue samples, which require a digestion step, over liquid samples. An expensive and potentially cumbersome aspect to extracting microbial DNA from human samples is the presence of host contamination, or human DNA within the isolated DNA. If host DNA is not removed before the sample is sequenced, a large portion of sequencing reads will be from this host DNA and not the microbial DNA. To increase the amount of microbial DNA in the sample prior to sequencing, microbial enrichment options are available to decrease the presence of host DNA and, therefore, increase the proportion of microbial DNA.

Samples taken from tissues and swabs tend to be oversaturated with the individual's own human DNA (>90% human genome-aligned reads for swab samples [50]). This oversaturation can obscure the infecting microbe(s) and increase the likelihood of false negatives. There are two general ways to remove host DNA from a sample—either depletion of host cells or enrichment of microbial cells [51,52]. While these techniques sound similar, the approaches are different. Depletion of human DNA is completed pre-extraction (Figure 2b) by a selective lysis and degradation of the eukaryotic cells followed by purification of the sample to remove the degraded cells. This process leaves the intact bacterial cells behind to be extracted. Recent work has been able to remove as much as 99% of human DNA from lung tissue samples [53].

Alternatively, enrichment of microbial cells is completed post-extraction (Figure 2b) by a methylation selection, where CpG-methylated host DNA is selectively bound and the microbial DNA is eluted [51]. For example, a 50-fold decrease in reads aligning to host genomes can be achieved using this method [51]. A review of clinical metagenomics studies from 2015-2019 show host-DNA depletion and microbial enrichment used in similar proportion across studies (Figure 4a). Both methods are capable of removing microbes from the sample during this step; host depletion will lyse fungal and protist cells and may also lyse bacterial cells, and not all microbial cells are methylation free. Therefore, using host depletion techniques prior to application of metagenomics methods may prevent the detection of parasitic protists, such as those that cause malaria or giardiasis. Fluids, as well, tend to be oversaturated with human DNA (>90% saliva metagenomic reads mapped to human genome [50]) but are also severely diluted due to their nature, which results in extremely low yields of microbial DNA within the sample. To circumvent this, fluids have been mostly studied by metataxonomic methods (i.e., 16S rRNA amplicon sequencing). However, efforts are being made to adapt metagenomics methods and microbial enrichment techniques to study these samples. For example, a microbial enrichment method increased the number of mapped reads to the *Neisseria gonorrhoeae* genome from 1% to 43% in clinical samples of urine [54].

The most common source of starting material for human metagenomic studies is fecal samples, which are often used as a proxy for studying the gut microbiome. Unlike other sample sources, fecal samples have less host DNA contamination[55] (<10% metagenomic reads mapped to the human genome[50]) and depletion of host DNA is not needed (Figure 4b). Where other samples, such as tissues and swabs which are often oversaturated with host-DNA, usually undergo either host-DNA depletion or microbial enrichment (Figure 4b). In a recent clinical study which used fecal samples to study fecal transplant in children with *Clostridioides difficile*, there was no host depletion step and 98.26% of all reads mapped to bacterial genomes while 1.66% of all reads mapped to viruses and archaea[56]. A researcher may not want to remove all host DNA, however, as a high level of host DNA in some samples (e.g., feces) is an indicator of various diseases such as Inflammatory Bowel Disease (IBD) and *Clostridium difficile* infection[57].

Standardizing procedures have been developed to minimize risk of contamination during processing samples for metagenomic analyses, but contamination may still occur and can cause problems for downstream inferences. Possible sources of contamination include sample to sample contamination [58], in which samples are mixed together; and reagent contamination, in which reagents can introduce microbes into the sample. Contamination can be especially detrimental for already low quantity microbial input samples and can affect study conclusions [59,60]. For example, a few lung samples were found to be dominated by contaminated microbes rather than microbes inherently present in the lung samples [61]. While an in-depth evaluation is out of the scope for this review, there exists a variety of reviews related to this topic [59,62,63].

Subsection: *Sequencing technologies*

Prior to the actual sequencing on NGS instruments, samples need to be processed through library preparation, which is the first step in NGS and specific to each sequencing platform. While library preparation is a standardized workflow, a major aspect that can affect sequencing quality, quantity, and downstream analyses is the amount of input microbial DNA in the sample (also referred to as "bacterial/microbial load"). While host-depletion or microbial enrichment options are available (addressed in previous subsection), the input sample needs to contain a minimal amount of microbial DNA, which is often dependent on sample type and extraction protocol. Otherwise, sequencing depth and coverage is wasted on non-microbial or host genomes. Most library preparations specify a minimal DNA input level, but experience and expertise in NGS and metagenomics is needed to understand and process samples for NGS.

Sequencing depth, the total number of sample reads that are produced by a sequencing machine, and sequencing coverage, the number of overlapping sample reads that cover a given nucleotide of a reference genome, are both important values to consider when processing a patient sample. In samples where host-DNA predominates (i.e., low microbial load in input sample), reduction in sequencing depth decreases both the number of microbes that are detected and the overall sensitivity [64]. Errors in nucleotide sequencing vary by technology and machine, but can be partially curtailed by increasing coverage, which can be achieved with increasing depth. A review of the current clinical metagenomic studies in the last five years show researchers have used sequencing depth as low as .0002 and as high as 9.75 Gbp per sample (Supplementary Table 1). Input microbial DNA amount can be used to determine the desired sequencing depth and coverage needed during sequencing.

Sequencing is increasingly performed using NGS technologies, which quickly yield reads at a rate that is orders of magnitude higher per run, and at a fraction of the cost per base, when compared to the Sanger method [16,65]. The sequencing market is currently dominated by Illumina (San Diego, California) platforms, which use short-reads (≤600 bp) and carry the lowest cost per Gigabase pair (Gbp) (Figure 5a; Table 1). Pacific Bioscience (PacBio) (San Francisco, California) and Oxford Nanopore Technologies (Oxford Science Park, UK) have developed platforms capable of sequencing long-reads (>10,000 – 2.2M bp). While these long-read platforms cost more, and have a higher error rate and lower throughput when compared to Illumina's short-reads [16] (Table 1, Figure 5b, 5c), they remain an improvement on Sanger with their drastically lower costs per Gbp (Table 1). The choice of which technology to use is dependent upon the research aims (Table 2), however, most clinical metagenomics studies use Illumina platform (Figure 5d) which sequences short reads. Short-reads tend to catch a larger number of individual microbial genomes than long-reads; however, short-reads cover fewer base pairs of those genomes than long-reads do (Figure 5e, 5f). Short read approaches typically have a lower error rate and cost. There are, however, additional benefits to the other platforms. Nanopore's minION, for example, is small and compact, allowing it to be used, with slightly altered workflows, anywhere from a forest to the microgravity of space [66]. PacBio and Nanopore can also be used to detect some methylation of DNA in epigenetics [67].

These technologies are continually improving. Multiple times a year, Illumina alone will upgrade their sequencing techniques on its current machines or produce an entirely new machine with enhanced technology and higher throughputs. Clinicians interested in integrating NGS sequencing technology must be wary that such updates would require revalidation of any diagnostic laboratory workflow using those machines.

Subsection: ***The computational challenges and limitations of metagenomics methods***

NGS technologies are capable of generating today's "big data" sets across large-scale clinical cohorts. Large data sets generated by metagenomics methods require the use of sophisticated bioinformatics algorithms that are capable of differentiating technical noise from biological signals in the data [68] and accurately assessing the metagenomic content of a given sample [69]. However, several challenges must be addressed in order to leverage the full potential of metagenomic computational techniques in the clinical setting, including the following: lack of standardization in bioinformatics techniques, widely varying performance of such algorithms, and the lack of comprehensive reference databases (Figure 6).

Developing an effective clinical diagnostic technique capable of properly informing medical decision-making requires specification of standardized, detailed, and replicable procedures [70,71]. This degree of detail is in contrast with academic research, where metagenomic computational techniques and pipelines are frequently published and distributed without standardized workflows, parameter settings, or input/output formats [72,73]. As a result, metagenomic bioinformatics pipelines are frequently updated and evolve as new algorithms and tools are published [74]. The ever increasing rate of newly developed computational methods and sequencing technologies causes the few existing published standardized procedures for metagenomic analysis (such as the Human Microbiome Project's "Manual of Procedures" [75]) to quickly become irrelevant. While some marker gene approaches, such as those using 16S rRNA, have begun to be standardized [76], there is less standardization and consensus for whole genome shotgun (WGS) methods [77]. Only recently have large-scale, unbiased assessments of individual WGS metagenomic tools been performed [72,73,78,79], with little in the way of standardizing entire computational pipelines. The academic software development community has made some effort to facilitate the ease of incorporating and assessing the performance of new methods in computational pipelines [80–83], but these efforts have not yet been widely adopted. This general lack of standardization results in many bioinformatics pipelines being created to analyze specific data types drawn from measurements of specific biological samples using specific sequencing technology. This "bespoke" approach to software development limits potentially wide adoption in the clinical setting where these exact conditions may not exist.

Today's field of metagenomics research is dominated by a wide range of bioinformatics tools that have been published with very different performance characteristics. As a rapidly developing field, metagenomic bioinformatic tools, especially those in WGS metagenomics, still have room for improvement [72,73,78,79]. One example can be found in metagenomic assembly — the process of building longer sequences (i.e., contigs or scaffolds) from the shorter reads that are output from a sequencing machine, which can then be used for further analysis such as assessing the gene content of a metagenome. One study shows that current assemblers struggle to resolve individual strains from a metagenomic sample [72,73], a task important for detecting pathogens. Taxonomic profiling, detecting the presence and relative abundance of microbial taxa in a given sample, is another common computational technique for which there is little guidance on how to select the best tool. The consensus among all large-scale benchmarking studies is that selecting a single "best tool" is often not possible or straightforward [72,73,78,79]. This difficulty is due to some tools excelling at maximizing different specific metrics, including: sensitivity (of the total microbial taxa actually present in a sample, maximizing the percent detected [true positives]), specificity (of the total microbial taxa absent in a sample, maximizing the percent not detected [true negatives]), and correct prediction of the relative abundance of taxa. No current analytic technique excels at all three (or even two) of these metrics [72,73] (Figure 7a, 7b). Expert-level proficiency is often required in order to select the best possible tool for a given clinical application, and many medical institutions now tasked with metagenomics research lack access to computational expertise. A review of the clinical metagenomics literature of the last five years showed that a variety of bioinformatics tools are in use, with one of the most popular being MetaPhlAn, which is used in 55% of selected clinical metagenomic studies (Supplementary Table 1; Figure 7c) even though other tools may perform better[84]. In addition, metagenomic computational tools struggle with poor performance and lack of high quality, comprehensive databases when asked to analyze samples at a resolution finer than the taxonomic level of genus [72,73]. These complications are an unfortunate limitation when information at the level of species or strain is required in clinical settings.

Broad application of metagenomic computational techniques are presently limited in the clinical setting due to the lack of comprehensive reference databases [47,48,74,85]. Most metagenomic tools use a reference database comprised of information concerning microbial organisms in order to identify material from samples—including whole genome sequence databases [86], databases of gene families [87], and databases of taxonomic relationships [88,89] (we refer to such information here collectively as "reference databases"). Currently available reference databases can be extremely variable in composition and quality, and no single database represents the totality of existing information. For example, a recent analysis of fungal reference genome databases shows a greater than 30% discrepancy [38] at the species level between different reference databases. Similar discrepancies have been found in protein orthology databases[90] and in taxonomic databases [12,91]. Furthermore, reference databases continue to experience unprecedented growth in size [92,93]. Some computational tools fail to utilize up-to-date reference databases due to the time-consuming [94] or difficult-to-implement nature of post-processing procedures [95]. The use of biased, incomplete, or outdated reference databases can further negatively influence metagenomic computational algorithm performance. Expert-level proficiency is often required to select the reference database(s) that is most appropriate, up-to-date, and correct. Furthermore, the computational skills required to apply and interpret metagenomics methods are not included in most graduate-level biological or biomedical curricula [96–98].

Subsection: ***Statistical challenges and limitations of metagenomics methods***

In addition to these computational and database challenges, there are few – if any – standardized statistical methods to analyze the output of the many different currently available metagenomic tools. For example, when seeking to test for differences in microbial composition between two or more populations, a variety of methods are currently used, including PERMANOVA [99–104], analysis of similarities (ANOSIM) [105–110], kernel machine regression [111–117], Dirichlet-multinomial distribution hypothesis tests [118–121], and sparse partial least squares discriminant analysis [122–126], to name a few. Such a variety of statistical methods extend to other biological questions [127].

While "best practices" do exist for microbiome analysis [77,128], even these mention the variety of available statistical methods available without recommending a standard methodological procedure. Indeed, existing computational packages intended for end-to-end analysis of microbiome data (e.g., mothur [129], QIIME2 [130], and Anvi'o [131]) include many statistical methods to be applied but require the user to determine when to apply these and if they are appropriate to use for their analysis. Unfortunately, this lack of clarity about ideal application of statistical methods in the face of many different study designs, biological questions, and the compositional nature of metagenomic profiles [132–134] can significantly affect the outcome and reproducibility [135–139] of study results, and is not conducive to clinical diagnostics. Unsurprisingly, this lack of clarity has driven researchers to base many clinical metagenomics studies solely on reports generated by the particular tool or pipeline utilized without additional statistical analysis (i.e., using only presence/absence data as reported by the particular tool in the study) [53,140–142].

**The hurdle of validation**

Culturing methods are a standard diagnostic method in clinical labs, but metagenomics methods are new. Data generation and data analysis technologies associated with metagenomics are continually changing, and considerable lack of standards exist among bioinformatics analytical software. A clinical lab implementing metagenomics methods must pay careful attention to the amount of time and financial resources required to continually revalidate metagenomics methods as the method is improved upon.

Laboratory Developed Tests (LDTs) are defined as diagnostic tests that individual clinical labs have developed and validated. LDTs are performed exclusively 'in-house' (i.e., not commercially distributed outside individual labs). These clinical labs and tests are certified, overseen, and regulated by Centers for Medicare and Medicaid Services (CMS), as stated by CLIA, and not necessarily the Food and Drug Administration (FDA) [143,144]; although the nature and scope of the FDA's role in certification is currently a topic of debate. LDTs, therefore, may not be FDA approved or validated but are required by CLIA to be validated by the lab itself and provide CMS inspectors with data that assures the accuracy and reproducibility of such a test. LDTs have been used to quickly respond to public health threats such as H1N1, SARS, and COVID-19 [143,144].

In recent years, the FDA has claimed authority to regulate a subclass of LDTs: In Vitro Diagnostic Multivariate Index Assays (IVDMIAs) [143–145], which includes genetic testing. As of this publication, there are no FDA validated metagenomic NGS diagnostic tests; however, clinical labs certified by the CMS under CLIA exist that offer mNGS testing on clinical samples [146]. These differ from *in vitro* diagnostic (IVD) test kits that are developed by manufacturers and commercially sold to clinical labs. The validation process for LDTs testing is complex and rigorous, needing to meet the requirements of the CMS through the CLIA, state regulations (which vary by state), American College of Medical Genetics and Genomics (ACMG) guidelines, and potentially FDA regulations.

If any stage of a workflow is altered, either because the sequencing technology has changed or if a single part of the bioinformatics pipeline is updated, the entire method must be revalidated to show that the sensitivity and specificity have not been negatively impacted before use in a clinical setting [147]. As developers and engineers are working to improve these technologies and software, there can be multiple changes made throughout a single calendar year. This culmination of changes imposes nontrivial challenges to any clinical lab trying to implement NGS technology and sequence analysis into the lab as they must provide data of the highest quality possible (see review [148]).

**Leveraging metagenomics approaches for clinical laboratories**

There currently exists great potential for leveraging metagenomics methods to develop tools capable of detecting microbiota associated with human disease, infectious disease, and foodborne illness. The microbiome has been proven to be an effective biomarker for conditions such as colorectal cancer [149–151], inflammatory bowel disease (IBD) [152,153], and various metabolic syndromes [154–156]. Moreover, metagenomics can potentially be leveraged to produce applications that can be used to determine the risk of infection during treatment procedures and can be used to develop pre-operative procedures to reduce risk of complications. In some cases, microbiome analysis may provide better insights into the etiology of disease than established diagnostic methods (e.g., blood sample analysis).

The strategy of linking metagenomics with other metadata and variables, like diet, has strong potential beyond gastrointestinal disorders. In addition to diagnosis of a known disease, metagenomics provides a unique opportunity to explore diseases of unknown etiology, which can be particularly helpful in the face of emerging, zoonotic, or rare infectious agents.

Furthermore, metagenomics has provided the opportunity to discover previously unknown and unstudied classes of viruses, which may have currently undiscovered roles in disease exhibited by outbreaks of H1N1, SARS, and COVID-19 [157]. Metagenomics allows the opportunity to study the interactions of different microorganisms (e.g., bacteria, virus, eukaryotic microbes) that produce disease phenotypes. This is, in fact, a paradigm shift from the last century of detecting/diagnosing '*the*' organism causing disease. Metagenomic techniques allow researchers to characterize the entire microbiome, including the relative contributions of individual components and interaction effects that contribute to disease phenotypes.

Subsection: ***Pathogen detection and treatment***

Metagenomics-based studies also have potential to revolutionize the way well-understood bacterial infections are identified and personalize treatment plans. When a patient presents with bacterial infections, the clinician typically prescribes broad spectrum antibiotics to rapidly begin treatment. A more specific diagnosis would allow clinicals to prescribe for the patient personalized antibiotics, but labs traditionally use time-consuming culture-based methods to identify the infecting agent. Culture-based methods can take multiple days, during which time the broad-spectrum antibiotics can harm the natural microbiome of the patient, worsen symptoms, or have no effect on the infection – potentially allowing it to continue spreading in the patient's body and even gain resistance to the antibiotics. A culture-independent technique using diagnostic PCR, used to amplify the DNA of specific microbes known to cause certain diseases, can fail to identify the harmful biota because of incorrect primer binding sites or overshadowing of minority taxa.

Metagenomics methods can provide clinically actionable results in hours (not days), and, with the use of amplification techniques of untargeted sequences, a clinician can deliver to the patient treatment personalized for the composition of the infection. These amplification techniques do incur biases in the resulting sequences, and informed caution should be used when utilizing them [158]. Further, application of metagenomics methods to polymicrobial infections cannot yet replace convenutal methods, but has strong potential as a diagnostic tool for both monomicrobial and polymicrobial infections [140]. Ultimately, metagenomics can help a clinician avoid prescribing for the patient potentially harmful broad-spectrum antibiotic treatment – ultimately limiting the evolution of new antibiotic resistant bacterial strains.

Metagenomics studies also show promise in the development of strain identification for *Mycobacterium tuberculosis* (TB), which has an emergent strain showing drug resistance to isoniazid and rifampicin, called MDR-TB. Currently available treatment plans for additional strains – resistant to fluoroquinolones and other antibiotics, called XDR-TB – carry only a 19% favorable outcome for the patient [81]. Analysis of globally reported cases of TB show 4% of cases are rifampicin resistant [159]. Drug resistance for cases of TB are traditionally determined using culture methods, whereby the infecting strain is grown in the presence of various antibiotics. Clinics currently use a method leveraging PCR - instead of culture – that can detect cases of TB on site and identify rifampicin resistance within two hours [160]. The results of this test can determine whether or not a patient should receive treatment for MDR-TB or the standard TB treatment. Although reliable for fully evolved MDR-TB strains, recent genetic analyses of the evolution of TB drug resistance have shown that the development of resistance to isoniazid occurs before the development of resistance to rifampicin [161]. Patients infected with TB at earlier stages of drug resistant evolution, resistant to isoniazid but not rifampicin, would not be identified as MDR-TB by the current methods. The treatment of these cases with the standard TB treatment allows for the further evolution and increased prevalence of antibiotic resistance into MDR-TB. Metagenomics methods, if successfully leveraged, could provide access to the relevant parts of the genome that inform antibiotic resistance and allow for more personalized treatment.

*M. tuberculosis* is not the only bacterium causing demand for more precise monitoring techniques. Harmful bacteria found in farm animals and meat produced for human consumption have developed a resistance to Tigecycline (an antibiotic of last resort to treat severe infections). The current cause of this resistance is associated with the evolution of two genes named tet(X3) and tet(X4), which inactivate all tetracyclines (the classification of antibiotics under which Tigecycline falls). Even newly FDA-approved antibiotics, including eravacycline and omadacycline, are ineffective at treating these bacteria. These two genes are found on plasmids and can therefore be copied and given to other bacteria through horizontal gene transfer. The spread of resistant bacteria and their genes have not been fully quantified; however, resistant strains have already been found in hospital patients in China [162]. Metagenomics methods have the ability to quickly determine the resistance of a bacterial strain before the administration of potentially ineffective drugs – a developmental key to improving health outcomes of thousands of individuals affected by antibiotic-resistant strains of harmful bacteria.

Encephalitis is an inflammation of the brain caused by an infection that, if left untreated, can result in serious disability and carries a 30% mortality rate. Currently, the instigating microbe in more than 33% of encephalitis cases are unidentified, and available reference databases for encephalitis are incomplete. Identification of the bacterial agent is typically performed using diagnosis PCR, which only amplifies the DNA of microbes already known to cause encephalitis. Metagenomics methods are capable of identifying the cause of encephalitis, and in the process, identify strains and organisms not previously associated with the condition. Therefore, metagenomics methods are capable of identifying malicious microbes (even those never associated with harmful infections) [163].

Subsection: ***Microbiome as a biomarker for disease***

Microbiomes associated with the human body are sensitive to changes in diet [164–168], medications [169–171], nutritional supplements [172], environment [173], and health conditions [174], so microbiome composition can therefore be used as a predictor of illness. Metagenomic analyses have established links between altered microbiome composition and disease, which often show a decrease in microbial diversity [175–181] . A few notable examples include establishing links between gluconeogenesis, putrefaction, and fermentation pathways in microbiota with colorectal cancer (CRC)—a common cause of mortality across the globe [182]; showing strong potential for the diagnosis of inflammatory bowel disease (IBD), with the capacity to predict patient response to treatment and the chances of post-treatment relapse; and identifying bacterial commonalities between IBD and other inflammatory disorders including psoriasis, psoriatic arthritis, spondyloarthritis, and rheumatoid arthritis, thus suggesting a common gut dysbiosis among these disorders [175–178]. Current diagnostic techniques for IBD and irritable bowel syndrome (IBS) often fail to distinguish between the two conditions, since they present similar symptomology, but metagenomic studies have led to the identification of key bacterial species and microbial composition in the intestines of individuals with possible links to both IBD and IBS conditions [152]. Besides fecal samples, other sample types have been shown to be useful in identifying microbiome biomarkers [181]; a number of tumor-associated microbiota have been linked to specific oral microbiota, suggesting that oral samples may provide an even easier sample site when compared to the gastrointestinal tract [183]. Studies focusing on the application of microbiome-based treatment have led to trial uses of microbiome transplants as treatment plans for various infections and conditions, and have the potential to replace standard antibacterial treatments that are quickly becoming ineffective [74,184–186].

Subsection: ***Microbiome as a dictator of personalized treatment***

The microbiome has a strong effect on an individual's ability to interact with, excrete, or metabolize food and medication. Assessing the impact that the microbiome has on the activity of pharmaceutical drugs will become essential for clinicians to develop personalized therapies for patients. Understanding the link between an individual's microbiome and metabolism, weight gain, and metabolic diseases can also be used in personalized medicine. Specifically, metagenomics-based approaches show promise in predicting an individual's probability of developing melanoma and IBD, as well as determining the amount of a medication that will actually be processed in a patients' body (bioavailability). Metagenomics has a clear role in potential development of microbiome-based treatment – including treating individuals with unhealthy gut microbiota, obesity, and diabetes. Protocols for this precision metagenomics have already been proposed [187].

One of the most notable and long-standing treatments regarding microbiome is Fecal Microbiota Transplant (FMT). FMT is considered a viable and FDA-approved treatment for *Clostridioides difficile* infection (CDI), which drastically decreases the fecal bacterial diversity in an affected person. While FMT is considered an effective treatment in adults, the adverse effects are not well known or characterized; recently, the death of a patient caused the FDA to issue a safety alert regarding the transfer of multidrug-resistant organisms by FMT [188]. This notification and a recent study involving FMTs in children [189] reinforce the importance of screening the donor with metagenomics methods to identify harmful or drug-resistant organisms prior to administering a FMT to the patient. The efficacy of this treatment strategy for more complex microbiota-associated gastrointestinal disorders remains to be tested, but trials are currently ongoing to test the application of FMT in IBD, IBS, and obesity [190]. Additionally, FMT has also been considered as a treatment for neuropsychiatric disorders, including autism, in which patients treated with FMT experienced improved symptoms [190,191].

Microbes (such as those in the phylum Bacteroidetes) have even been shown to interfere with the liver's decomposition of widely used drugs, such as acetaminophen, leading to the increased formation of more toxic and carcinogenic compounds [10]. Additionally, a decrease in Bacteroidetes abundance has been seen in women who have indicators of blood glucose control issues [192–194]. Aside from modulating drug activity, the most important potential applications can be found in the administration of cancer treatments. The presence of certain microbes can increase or decrease the toxicity or efficacy of treatment [11,195]. Having an outline of the microbes present inside a patient can inform doctors on the appropriate course of drugs.

Application of clinical metagenomics to phenotypes is not limited to disease, and can also be applied to phenotypes associated with mental health issues. Human genome analysis has been frequently used when studying the development of schizophrenia [196]; however, in a recent 2018 study utilizing blood samples, patients with schizophrenia are shown to have higher microbiome diversity than those with amyotrophic lateral sclerosis and bipolar disorder or healthy controls [197]. The gut microbiome of schizophrenia patients, too, differed from gut microbiomes of healthy control groups [198]. There is evidence that the gut microbiome is capable of affecting behavior associated with schizophrenia [198], and the connection between host-associated microbiota and mental health extend beyond schizophrenia to phenotypes for anxiety and trauma-related disorders [199].

**Discussion**

Metagenomics methods have a wide range of utility and applications. These include, but are not limited to, infection detection, surveillance and tracking of outbreaks, antimicrobial resistance prediction and strain resolution, and disease prognosis. Overcoming the current limitations of metagenomics methods is necessary not only for the field of microbiology to progress, but also to acquire clinical validation. Metagenomic methods must be reproducible, reliable, and accurate to be clinically actionable [200]; currently, most metagenomic methods and tools are not. As of publication, only three validated clinical tools exist that are based on metagenomics methods. The first is the use of cerebrospinal fluid to identify the infectious agent in encephalitis and meningitis cases, which is performed by the Clinical Microbiology Lab at University of California, San Francisco. The second is the use of bronchoalveolar lavage fluid for lower respiratory tract infections, completed by IDbyDNA. Finally, identification of microbial cell-free DNA from blood plasma is implemented by Karius Inc. All of these methods underwent rigorous scrutiny, and researchers have published papers providing evidence that their method is reproducible, reliable, and accurate [201–203].

Clinical validation also depends on financial interest in the medical research and development community. Insurance companies traditionally have not covered the costs of metagenomic tests for individuals who wish to use available novel methods. The withholding of financial coverage is due to a lack of sufficient data demonstrating that these methods can effectively improve patient outcomes. Currently, patients need to pay out of pocket for a test leveraging metagenomics analysis; the financial barrier results in low use of the test. There need to be larger studies and pilot programs that can demonstrate these methods effective, capable of significantly improving patient outcomes, lowering health costs, and reducing dependence on general antibiotics. More funding is needed for clinical metagenomics studies and towards the maintenance of pipelines for clinical use so that metagenomics can be effectively applied as a clinical tool. While clinical metagenomics has a number of obstacles to overcome, the potential benefit to the medical field promises to vastly outweigh the cost of development.

**Methods**

To generate Figure 1, we used [204] to count the number of publications by year. Keywords "metagenomics" and "clinical metagenomics" were used to search for papers. The keyword search was limited to the title and abstract only.

**Data availability**

The data used to generate the graphs for figure 1 and the bar charts in figure 5 are present in supplementary materials (Supplementary Tables 2, 3, and 5).

Code and data for generating figures 3, 4, 5d, and 7c can be found at

https://github.com/LizChang/Review-Paper-Figures

**Figures and figure legends**

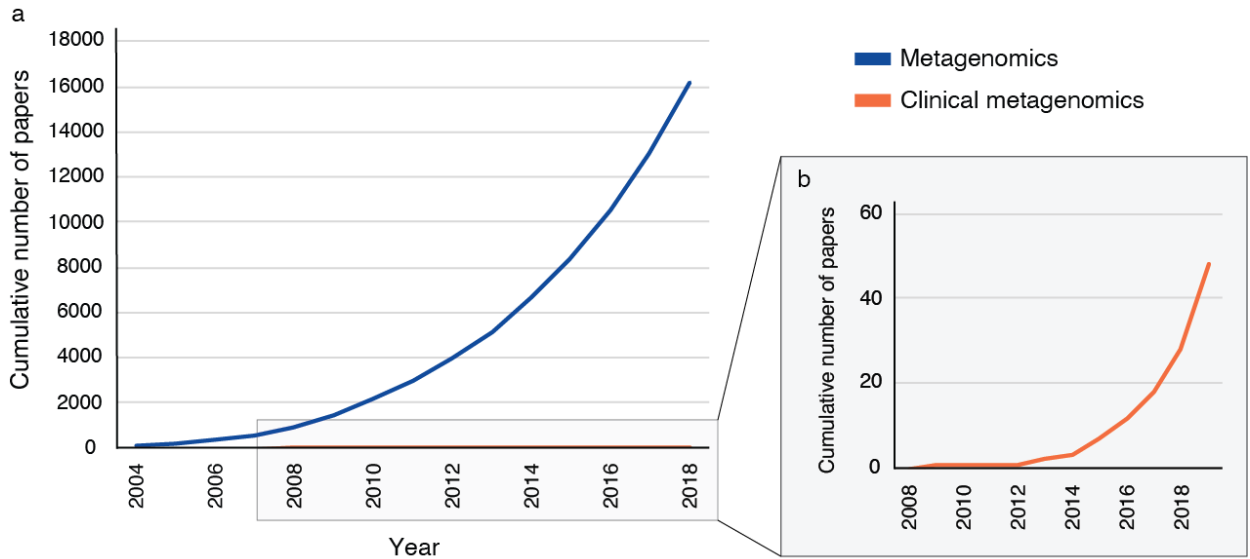

**Figure 1. Rise of metagenomics and trailing of clinical metagenomics** (a) The cumulative trend of papers published on metagenomics (blue) and clinical metagenomics (orange) from 2004 to 2019. (b) Enlarged graph of the cumulative trend of papers published, focusing specifically on the clinical metagenomics trend from 2008 to 2019. Note: data cumulated using [204].

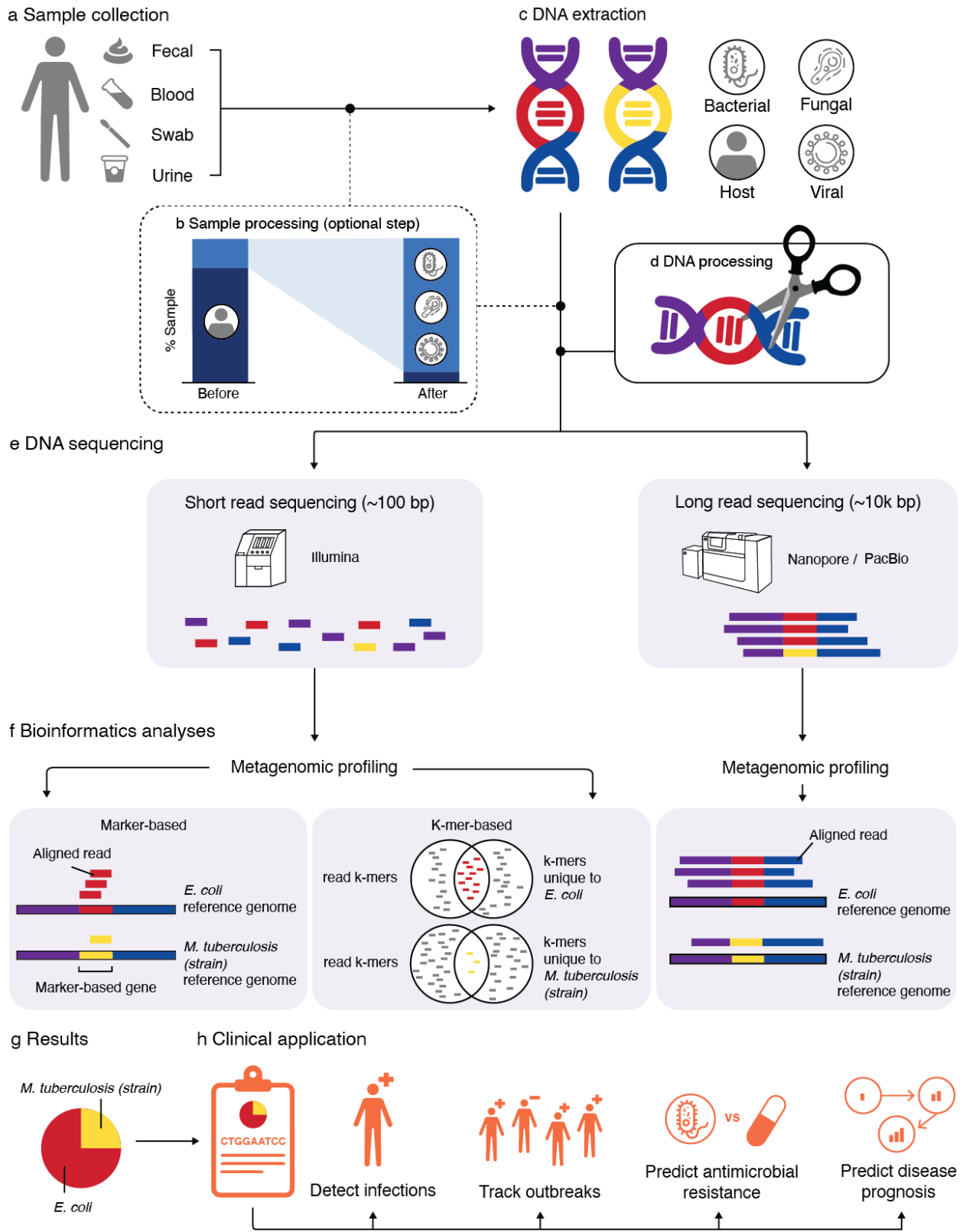

**Figure 2. Clinical metagenomics workflow.** (a) Sample collection. The first step of clinical metagenomics is sample collection from the patient. Samples take the form of any fluid (e.g., blood, urine), solid (e.g., feces), or alternate sampling method (e.g., swabs, biopsies) taken from the human body which can then be stored at varying temperatures or with a preservation solution to keep the nucleic acids intact. (b) Host-DNA depletion or microbial enrichment. Samples taken from living hosts may contain an overabundance of host DNA that may overshadow reads; often, host DNA can be mostly removed through host-DNA depletion (before DNA extraction) or microbial enrichment (after DNA extraction). These steps are optional. (c) DNA extraction. Genetic content of the sample is extracted from cells and isolated. (d) DNA processing. Extracted DNA is sheared and prepared for sequencing within a sequencing library. (e) DNA sequencing. The prepared DNA libraries are provided to a sequencing platform that determine the base pair pattern of the DNA. Five sequencing platforms are Illlumina, ThermoFisher (not shown) (Waltham, Massachusetts), BGI (not shown) (Cambridge, Massachusetts), Nanopore, and PacBio. The latter two generate long-reads (>1,000 bp) while the former three generate paired-end short-reads (<300bp) or single-end short-reads (≤600 bp). The acquired sequences are compiled into a computer file. (f) Bioinformatics analysis. Sequence compiled computer files become input for a metagenomics profiler (mapping shown). There are three types of mapping algorithms used to identify reads and measure organism relative abundance. The first is mapping-based profiling which maps the entire read to the entire reference. The second is marker-based profiling, which only maps the reads to parts of the genome that are unique to each species or strain and not to regions that are homologous. This step reduces the time it takes to map the reads. Finally, k-mer based profiling, which breaks up the reads and references into substrings and compares the substrings. (Binning methods are not discussed. See [84] for an evaluation of these different approaches) (g) Results. After the reads' identities and relative abundances have been found, bioinformatics analysis returns data on the microbial community. (h) Clinical application. Output of the computational algorithm is reviewed and compiled into a report used to inform clinical decisions. Clinical metagenomics can be used to detect infections, track the progress of outbreaks, predict antimicrobial resistance, and inform doctors of the best disease prognosis.

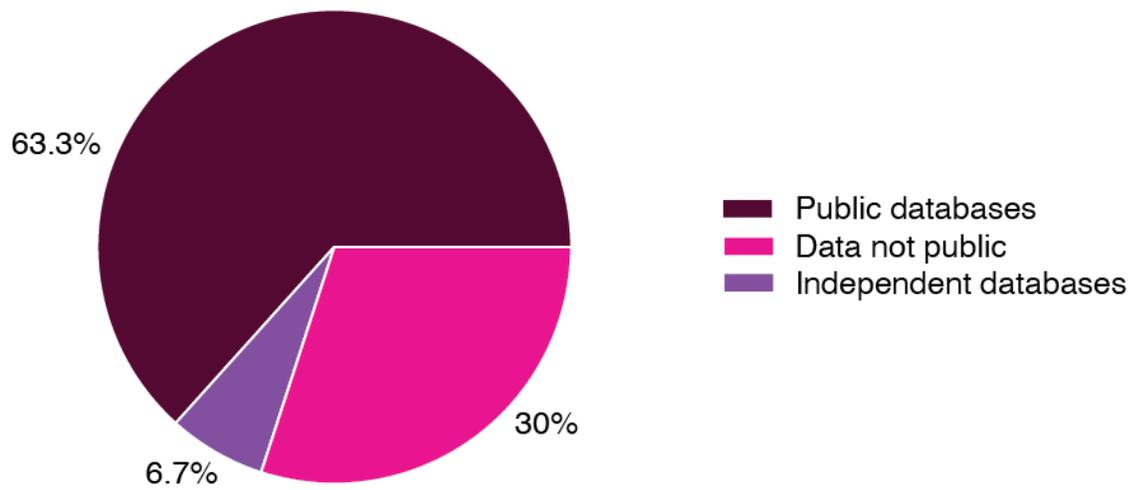

**Figure 3. Fate of raw data.** Percent of the 29 clinical metagenomics projects reviewed that has deposited raw metagenomics data in specialized public repositories, independent repositories, or not shared.

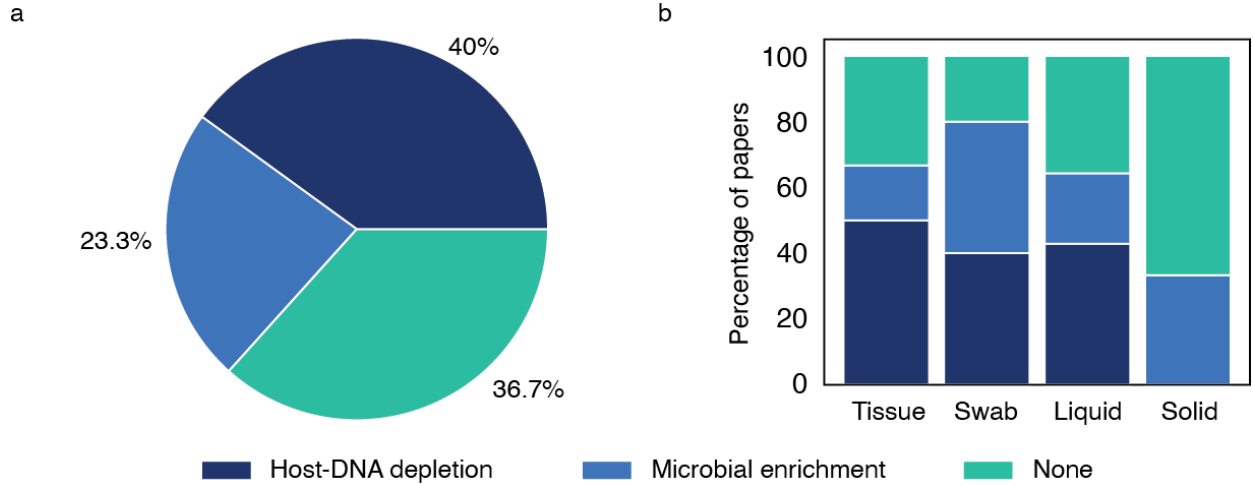

**Figure 4. Use of DNA processing statistics.** (a) Percent of the 29 clinical metagenomics projects reviewed that has utilized Microbial enrichment, host-DNA depletion, or neither while processing DNA for sequencing (b) The percentage of patient sample types that did or did not undergo microbial DNA isolation methods as a pre-sequencing step (Tissue, Swab, Liquid [e.g., urine, blood], and Solid [e.g., feces]).

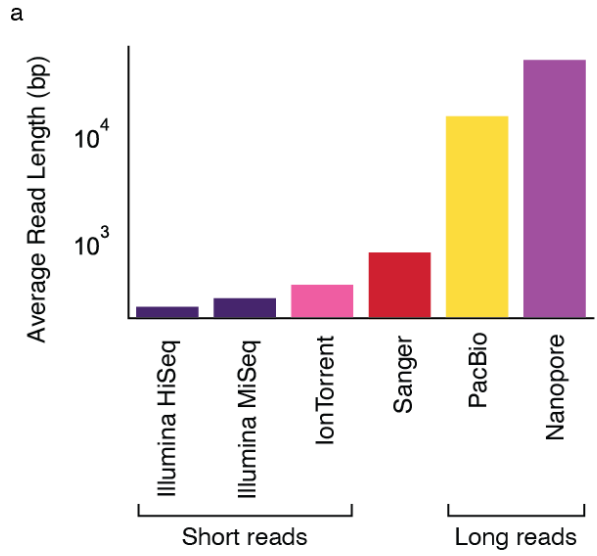
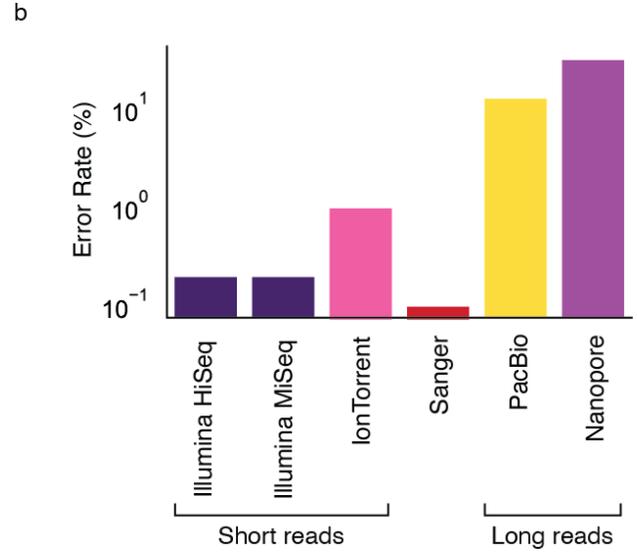
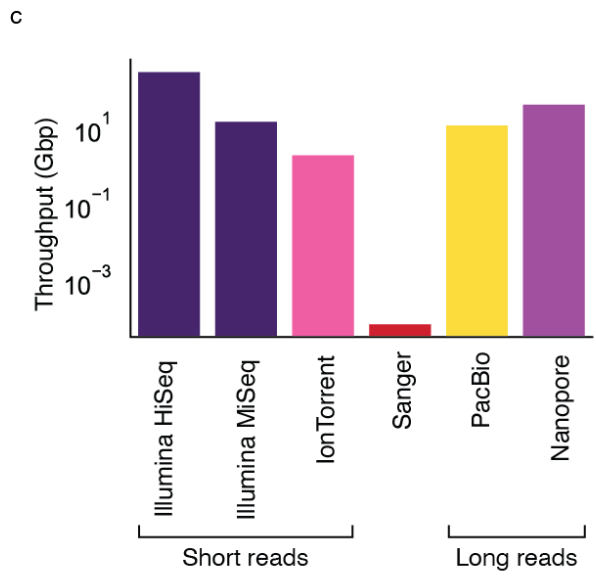
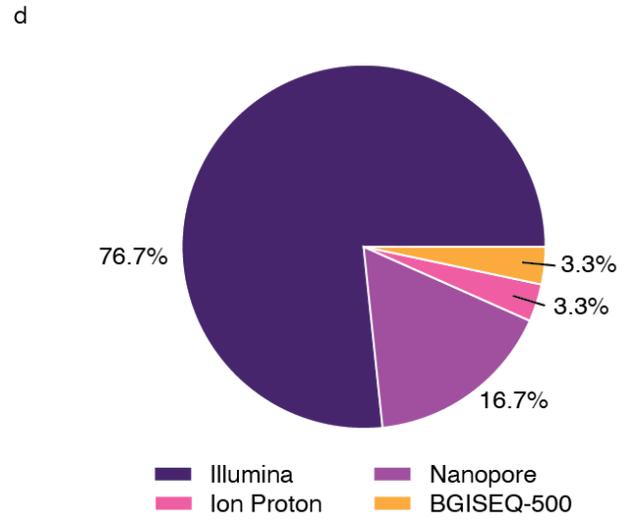
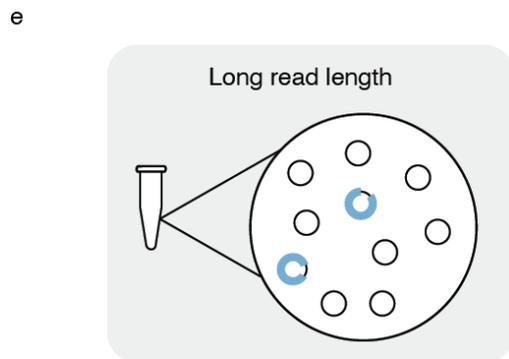
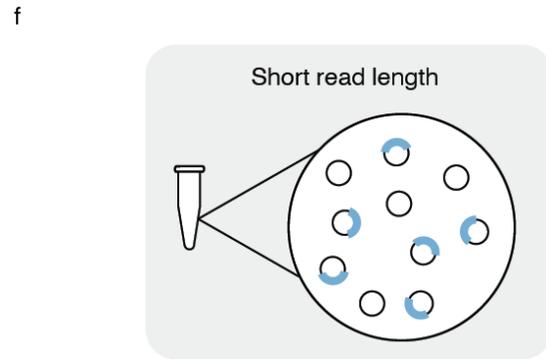

**Figure 5. Overview of Sequencing Technology.** Read length by sequencing technology (vertical axis log based) (data presented in Supplementary Table 2). (b) Error rate by sequencing technology (vertical axis log based) (data presented in Supplementary Table 3). (c) Throughput by Sequencing Technology (vertical axis log based). Illumina HiSeq data estimated based on HiSeq 2500 machine. Sanger data estimated using API Sanger 3500. Data for IonTorrent estimated using IonTorrent PGM. Data for Nanopore estimated based on MinION. Data for PacBio estimated based on Sequel II machine. Average read length and throughput data obtained from 'Developments in high throughput sequencing' [205] (data presented in Supplementary Table 3). (d) The percentage of clinical projects that used Illumina, Nanopore, Ion Proton, BGISEQ-500 to sequence sample DNA in 29 recent (2015 – 2019) clinical metagenomics studies. (e) Long reads. Long read sequencing obtains reads from few species, but read coverage is high — representing (in blue) almost the complete coverage of the species genome (black circles). (f) Short reads. Short read sequencing obtains reads from many species but reads only represent (in blue) a small fraction of the genome (in black) — representing only a small portion of the assembly of the species genome. Read coverage is represented in blue.

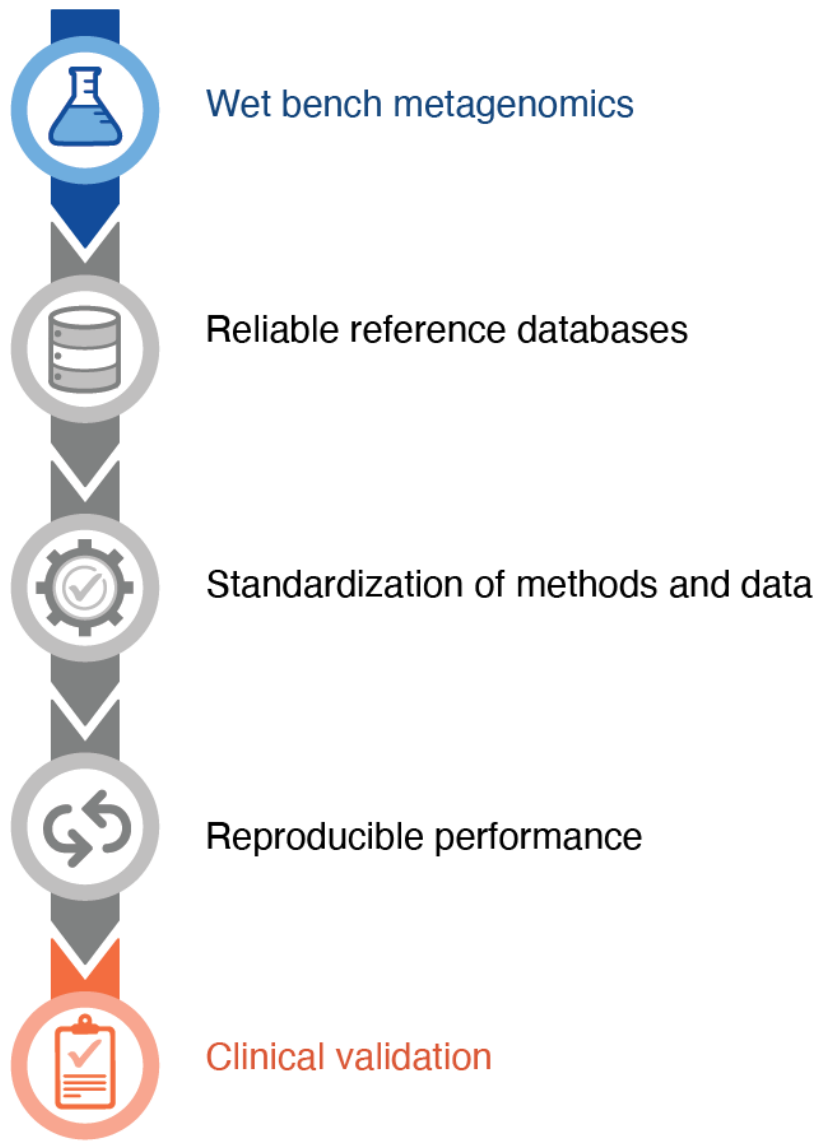

**Figure 6. Missing characteristics needed for clinical validation.** From the wet-lab bench, metagenomics methods have yet to develop reliable references databases, standardize methods and data, and perform in a manner that is reproducible. All these are required in order for the methods to become clinically validated.

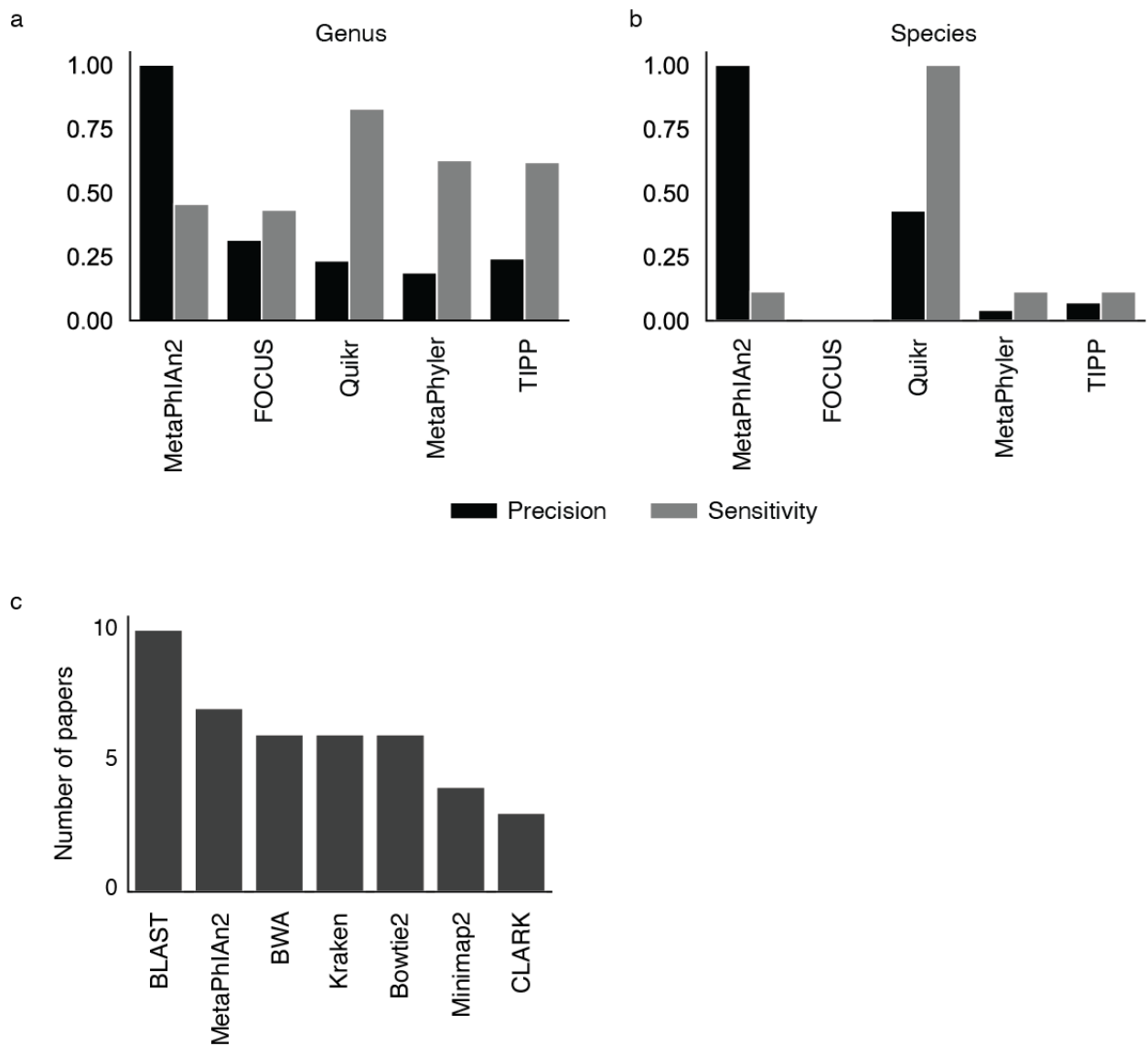

**Figure 7. Taxonomic profiler statistics.** (a) A side by side comparison of sensitivity and precision at the genus level. There is no profiling technology with both precision and sensitivity both above 0.50. Based on CAMI results[72]. (b) A side by side comparison of sensitivity and precision at the species level. There is no mapping technology with precision and sensitivity both above 0.50. Based on CAMI results[72]. (c) The most popular bioinformatics tools that are used among the 29 clinical metagenomics papers reviewed.

**Tables and table legends**

**Table 1. Comparison of available sequencing platforms.**

| Platform | Company | Error Rate | Average Read Length (bp) | Bases per run (gigabases) | Time spent per run (hrs) | Cost per Gb (US Dollars) | Release Year |
|---|---|---|---|---|---|---|---|
| ABI Sanger 3730xl | Applied Biosystems | 0.001 | 800 | 0.0000768 | 0.5-2.5 | $5,000,000 | 2002 |
| IonTorrent PGM | Ion Torrent | ~1% | 400 | 2 | 7.3 | $460 | 2013 |
| Illumina MiSeq | Solexa, Illumina | 0.20% | 300 | 15 | 21-56 | $110 | 2013 |
| Illumina HiSeq 2500 | Solexa, Illumina | 0.20% | 250 | 300 | ~40 | $45 | 2014 |
| PacBio RSII | Pacific Biosciences | ~13% | 13,500 | 12 | 2 | $600 | 2013 |
| Nanopore MinION | Oxford Nanopore | 32% | 9545 | 42 | ~0-50 | $750 | 2015 |

"Platform" is the sequencing machine referenced. "Company" is the current producer of the sequencing platform. "Error rate" is the percentage of the base pairs that are incorrectly read during sequencing. "Average read length" is the average length, in base pairs, of the reads generated by the sequencing platform when fed DNA. "Bases per run" is the throughput, in gigabases, that the machine can output every time it processes a sample of DNA. "Time spent per run" is the amount of time the sequencer takes to process a sample of DNA. "Cost per Gb" is the amount of money, in US currency, to produce one Gigabyte of sequencing data. "Release year" is the year the sequencing platform was released for scientific use. Average read length, Bases per run, and release year were pulled from 'Developments in high throughput sequencing' [205]. For ABI Sanger, Error Rate and Cost per Gb was acquired from a paper entitled 'Comparison of Next-Generation Sequencing Systems' [206].

**Table 2. Clinical applications and obstacles of applying metagenomic NGS (mNGS).**

| Clinical Application | Requirement for metagenomics analysis | Sequencing Technology | Databases |
|---|---|---|---|
| Infection detection | Highly- accurate sequencing methods | Illumina/IonTorrent PacBio/Nanopore | Comprehensive |
| Surveillance and tracking of outbreaks | Sample collection across large number of individuals | Illumina/IonTorrent 16S/18S/ITS | Sequencing Technology Specific |
| Rapid diagnostics | Speed (<4h) | Illumina (iSeq) PacBio/Nanopore | Sequencing Technology Specific |
| Antimicrobial resistance prediction, strain resolution | Low read error rate | PacBio/Nanopore | Customized for research task |
| Disease prognosis | Modeling dynamics | Illumina/IonTorrent PacBio/Nanopore 16S/18S/ITS | Custom |

Clinical applications of metagenomics methods are a few of the specific areas of clinics to which metagenomics methods can be applied. Requirement for metagenomics analysis covers the features of methods that are needed for metagenomics to be validated within the given clinical application. Sequencing technologies list the current NGS technologies that would be needed for the given clinical application. Illumina and IonTorrent represent short-read sequencers. PacBio and Nanopore represent long-read sequencers. 16S, 18S, and ITS represents amplicon sequencing. Amplicon sequencing is not metagenomics, but rather metataxonomy. A comprehensive database is one that that has a representative reference organism at the species level for every organism present in the sample.

**Journal title**

*Genome Biology*

**Article type**

Review

**Paper title**

Metagenomics for clinical diagnostics: technologies and informatics

**Authors**

Caitlin Loeffler[1,2], Keylie M. Gibson[3], Lana S. Martin[2], Yutong Chang[4], Jeremy Rotman[2], Ian V. Toma[5], Christopher E. Mason[6,7,8], Eleazar Eskin[1,9,10], Joseph P. Zackular[11], Keith A. Crandall[3,12], David Koslicki*[13,14,15], Serghei Mangul*[2]

**Supplementary Material**

**Supplementary Table 1. An overview of clinical metagenomics studies published between 2015-2019.**

| Reference | Host-DNA depletion | Sequencing technology | Sequencing depth (Gbp) | Tissue | Disease | Data public | Sample size (n) | Bioinformatics tools | Fresh or frozen samples |
|---|---|---|---|---|---|---|---|---|---|
| [1] | Yes | Illumina HiSeq 2500 | X | Bone and joint | Bone and joint infections | No | 47 | MetaPhlAn2 Kraken Bowtie2 | Frozen |
| [2] | Yes | Illumina MiSeq | 2 | Aortic valve | Infective endocarditis | Yes | 1 | CLARK MetaPhlAn2 mothur UBLAST | Frozen |
| [3] | No | Illumina MiSeq | 0.041 | Aortic valve | Infective endocarditis | Yes | 1 | CLARK USEARCH MetaPhlAn2 BWA BLAST | Frozen |
| [4] | No | Illumina MiSeq | N/A | Cerebrospinal fluid | Toscana virus | Yes[I] | 1 | TOSV SmaltAlign VirMet | Fresh |
| [5] | No | Illumina GAIIx | 0.07 | Sputum[*] | Polymicrobial infections | Yes | 1 | SMALT BLASTn | Frozen |
| [6] | Yes | Illumina NextSeq | 9.75 | Feces and rectal swabs | Klebsiella pneumoniae | Yes | 2 | Canu Pilon Snippy SPAdes Prokka IDBA-UD ABRicate[B] mlst[B] ESOM BLAST Kraken | Fresh |
| [7] | Yes | Illumina MiSeq | 0.0002 | Liver | Chronic hepatic brucelloma | Yes | 1 | Mothur Kraken BWA | Frozen |
| [8] | Yes | Illumina HiSeq 2000 and 2500 | 1 | Broncho-alveolar lavage | Chronic pneumonia | Yes | 1 | CLARK Kraken MetaPhlAn2 USEARCH UBLAST | Frozen |
| [9] | Yes | Illumina HiSeq | X | Cerebrospinal Fluid | Meningitis and encephalitis | No | 204 | MAFFT PhyML SURPI+ | Frozen |
| [10] | Yes | Nanopore MinION | N/A | Sputum, bronchoalveolar lavage (BAL) and endotracheal aspirates (ETAs) | Lower respiratory infections | Yes[I] | 40 | Porechop Minimap2 Canu BLAST WIMP ARMA | Frozen |
| [11] | No | Illumina HiSeq 2500 | 1.85 | Resected prostheses | Prosthetic joint infections | Yes | N/A | MetaPhlAn2 BioBloom BWA BBMap | Fresh |

"Ref" is the citation for the clinical metagenomics paper. "Host-DNA depletion" is yes/no do they process the sample to remove host DNA *before* sequencing. "Sequencing technology" is the sequencing technology used to sequence the sample. Tissue is the source of the sample from the subject. "Sequencing Depth" is given in Gbps. This is the throughput of each study, an 'X' is given when the data is not publicly available for analysis. "Disease" is the ailment that the study focuses on. "Data public" are the raw reads from the study publicly available (yes/no). "Sample size (n)" is the number of total participants in the study. "Bioinformatics tools" are the computer programs used to analyze the sequenced data. "Fresh or frozen samples" tells whether the samples taken were frozen before sequencing or used fresh.

[*]Sputum refers to saliva and mucus that is eliminated from the respiratory tract.
[B]BLAST based tool
[I]Data only available on non-SRA site

**Supplementary Table 2. Data used to generate Figure 5, panel a.**

| Technology | Average read length (bp) |
|---|---|
| Illumina HiSeq 2500 | 250 |
| Illumina MiSeq | 300 |
| IonTorrent PGM | 400 |
| ABI Sanger | 800 |
| Nanopore MinION | 9,545 |
| PacBio | 13500 |

**Supplementary Table 3. Data used to generate Figure 5, panel b.**

| Technology | Error Rate (%) |
|---|---|
| Illumina HiSeq 2500 | 0.20% |
| Illumina MiSeq | 0.20% |
| IonTorrent PGM | 1% |
| ABI Sanger | 0.00% |
| Nanopore MinION | 32% |
| PacBio | 13% |

**Supplementary Table 4. Data used to generate Figure 5, panel c.**

| Technology | Throughput (Gb) |
|---|---|
| Illumina HiSeq 2500 | 300 |
| Illumina MiSeq | 15 |
| IonTorrent PGM | 2 |
| ABI Sanger | 0.0000768 |
| Nanopore MinION | 42 |
| PacBio | 12 |